\documentclass[11pt,english]{article}
\usepackage[T1]{fontenc}
\usepackage[latin9]{inputenc}
\usepackage{geometry}
\usepackage{color}
\usepackage{float}
\usepackage{amsmath}
\usepackage{amssymb}
\usepackage{graphicx}
\usepackage{setspace}
\usepackage[authoryear]{natbib}
\makeatletter

\providecommand{\tabularnewline}{\\}

\date{}

\@ifundefined{showcaptionsetup}{}{%
 \PassOptionsToPackage{caption=false}{subfig}}
\usepackage{subfig}
\makeatother

\usepackage{babel}
\begin{document}

\title{Change Detection in a Dynamic Stream of Attributed Networks{\let\thefootnote\relax\footnotetext{\\ Mr. Reisi Gahrooei is a doctoral student in the department of industrial and systems engineering at Georgia Tech. His email address is mrg9@gatech.edu. \\ 
Dr. Paynabar is an Assistant Professor in the department of industrial and systems engineering at Georgia Tech. His email address is kamran.paynabar@isye.gatech.edu. He is the corresponding author. }}}

\author{Mostafa Reisi Gahrooei, Kamran Paynabar\\
Georgia Institute of Technology, Atlanta, GA 30332, USA}
\maketitle
\begin{abstract}
While anomaly detection in static networks has been extensively studied,
only recently, researchers have focused on dynamic networks. This
trend is mainly due to the capacity of dynamic networks in representing
complex physical, biological, cyber, and social systems. This paper
proposes a new methodology for modeling and monitoring of dynamic
attributed networks for quick detection of temporal changes in network
structures. In this methodology, the generalized linear model (GLM)
is used to model static attributed networks. This model is then combined
with a state transition equation to capture the dynamic behavior of
the system. Extended Kalman filter (EKF) is used as an online, recursive
inference procedure to predict and update network parameters over
time. In order to detect changes in the underlying mechanism of edge
formation, prediction residuals are monitored through an Exponentially
Weighted Moving Average (EWMA) control chart. The proposed modeling
and monitoring procedure is examined through simulations for attributed
binary and weighted networks. The email communication data from the
Enron corporation is used as a case study to show how the method can
be applied in real\textendash world problems.
\end{abstract}

\section*{Introduction}

The relationship of entities in a complex physical, biological, cyber,
and social system can mostly be captured by networks. As such, development
of mathematical models and analytical tools that can characterize
the interaction of network entities has attracted significant attention.
In early efforts, most research has focused on static modeling of
networks, in which either a single snapshot or aggregated historical
data of a system are used for modeling and analysis \citep{erdds1959random,frank1986markov,kim2012multiplicative,hoff2002latent}.\textbf{
}However, in reality, most networks represent time-varying systems
that exhibit intrinsic dynamic behavior. In other words, the underlying
structure of such networks slowly evolves/changes over time. As a
result, recent studies have focused on modeling and analysis of dynamic
networks. Examples of such studies include Erdos\textendash Renyi\textendash Gilbert
model, in which an edge or a group of edges are added to the network
over time with some fixed probability \citep{erdds1959random}, Barab�si\textendash Albert
model that employs preferential attachment to connect a new node to
an existing network \citep{barabasi1999emergence}, small\textendash world
model, in which shortcut edges are added to an initial regular network
with probability proportional to the distance between the nodes \citep{watts1998collective,kleinberg2002small},
and Markovian models that allow both vertices and edges to be added
and deleted over time \citep{xu2014dynamic,sarkar2005dynamic,hanneke2010discrete,goldenberg2010survey}.
The underlying assumption of these models is that the network slowly
evolves/changes over time without abrupt changes in their underlying
mechanism. In reality, however, the occurrence of shocks and abrupt
changes in dynamic networks is very common. For example, resignation
of a key employee or occurrence of a conflict in an organization may
cause a significant change in the professional network of employees.
Therefore, developing tools and techniques to detect such shocks is
critical for the analysis of dynamic networks. 

This paper focuses on combining dynamic network modeling of a complex
phenomenon with statistical process control (SPC) techniques for quick
detection of temporal changes in network structures. The objective
is twofold: first, to determine the underlying mechanism that governs
the edge formation of a dynamic network stream during a reference
(in-control) period; and, second, to identify time periods when edges
are generated from different mechanisms. In a corporation, for example
(as will be discussed in the case study section), analysis of employees
communications helps better understand the organizational structure
and interactions. At the same time, it would be useful to detect structural
changes in the communications network and find the corresponding root
causes such as a fraudulent action or a significant change in the
organizational structure. The main challenge in developing an effective
monitoring method for dynamic network streams is to capture and distinguish
the gradual change of a network stream resulting from the underlying
dynamic mechanism from abrupt changes caused by shocks to the system.
Otherwise, the false alarm rate of the monitoring procedure significantly
increases. \textcolor{black}{By a gradual change, we refer to any
form of autocorrelation (i.e. temporal correlation) in a network stream.
For example, in a colleagues\textendash and\textendash family network,
a seasonality behavior might be expected, as one might communicate
with colleagues during a week, and with family during the weekends.
Such a trend is part of the natural behavior of the network stream
and should not be detected as a change, rather it should be captured
in the model as a gradual trend. Ex}isting methods fail to address
this challenge, and therefore, they are not effective in monitoring
dynamic networks.

In the past decade, change detection in network streams has received
special attention\textbf{ }in the computer science community \citep{ranshous2015anomaly}
and various detection methods have been developed based on community
discovery and similarity measures \citep{duan2009community,papadimitriou2010web,koutra2013deltacon},
compression techniques \citep{hirose2009network}, tensor decomposition
\citep{sun2006DynamicTensor,sun2006windowTensor,koutra2012tensorsplat},
and probabilistic modeling \citep{Aggarwal2011}. However, these studies
lack a comprehensive statistical component that distinguishes significant
changes (assignable causes) from those that might naturally appear
due to random disturbances (random causes). Moreover, they do not
explicitly model the dynamic evolution of the system. Another group
of studies, mainly proposed by the statistics community, employ SPC
charts to identify changes. A comprehensive review of these methods
is given by \citet{woodall2016overview}. Many of these methods, however,
focus on monitoring of the network connectivity measures including
average degree, closeness, betweenness, and density to detect temporal
changes \citep{mcculloh2009detecting,mcculloh2011detecting}. These
measures can also be calculated for a fixed window scanning over a
graph (sub\textendash graphs) to obtain the so\textendash called scan
statistics \citep{marchette2012scanOnGraphs}. Monitoring scan statistics
\citep{priebe2005scan,neil2013scan,sparks2013spatio} can improve
the detection of local changes in network streams, but it is computationally
expensive.\textcolor{black}{{} Recently, \citet{azarnoush2016monitoring}
combined the statistical modeling of attributed (labeled) binary networks
with SPC to detect changes in network streams. In this study, they
showed that monitoring the connectivity measures can overlook particular
forms of change in network streams, and defined the connectivity of
two nodes as a function of node attributes (e.g. age, sex, education,
etc.) to model the network structure and to detect temporal changes.
One major drawback of all the aforementioned studies, including \citet{azarnoush2016monitoring},
is that they do not properly consider the dynamic evolution of network
streams. That is, these works extract the features or model the entire
network independent from the previous network snapshots and ignore
the potential autocorrelation in the network stream.} In order to
address this issue, this paper leverages network attributes to explicitly
model network dynamics and to separate it from abrupt changes. Many
real-world systems can be modeled using attributed (labeled) networks.
For example, in the Enron corpus \citep{perry2013statistical,xu2014dynamic},
each node represents an employee with an attribute denoting his/her
role in the company (i.e. president, vice\textendash president, CEO,
manager, etc.). In a social network, each user might have attributes
such as age, location, occupation, etc. These attributes can provide
an effective means for statistical modeling and monitoring of a network
stream.

The main goal of this paper is to propose a new modeling and change
detection methodology for attributed network streams that exhibit
intrinsic dynamic behavior. For this purpose, we integrate GLM used
for static modeling of attributed networks with state transition models
that capture the dynamic behavior of network streams. The integrated
model updated over time using the extended Kalman filter (EKF) along
with an exponentially weighted moving average (EWMA) control chart
creates a monitoring procedure for quick detection of abrupt changes
in network stre\textcolor{black}{ams. Our proposed method in this
paper extends monitoring approach in \citet{azarnoush2016monitoring}
in two ways. First, we extend their network modeling method for binary
edges to weighted networks through GLM. That is, our proposed method
is capable of modeling a variety of network streams wherein the edges
can be modeled by a distribution from the exponential family including
Bernoulli (e.g. binary networks), Poisson (e.g. weighted networks),
normal (e.g. flow network), etc. Second, we consider the potential
dynamic (autocorrelation) in the network stream by including a state
transition equation over the network parameters. \citet{azarnoush2016monitoring}
try to take this dependency into account by using a moving window
approach that combines the data of the current network snapshot with
a few previous snapshots when estimating the network parameters. This
approach, although simple, is only effective where the network dynamics
is very simple (e.g. linear trend). If used for more complex dependencies
(e.g. autoregressive) the moving window approach will lead to a large
false alarm rate (or less detection power) due to residual dynamics
not captured by this approach. For example, in a colleagues-and-family
network, a seasonality behavior might be expected, as one might communicate
with colleagues during a week, and with family during the weekends.
Modeling these networks independent from the previous ones (or even
with moving window) fails to capture such a trend and increases the
number of false alarms. To alleviate this issue, this paper includes
an explicit state transition model to capture the correlation between
the network snapshots.}

\textcolor{black}{It is noteworthy to mention that \citet{Wang2012}
used control charts to monitor the Kalman estimation of the time that
an order completes an stage in a supply chain network. In this work,
the supply chain network is fixed and does not change over time, and
hence their method cannot be used for modeling and monitoring of dynamic
structure of a network stream. In our work, however, we provide a
dynamic model to capture and separate both gradual trends and abrupt
changes in a stream of networks.}

The organization of the paper is as follows: We begin with an overview
of the proposed methodology along with the notations used in this
paper. Next, we describe a static model for attributed networks built
on the GLM, and extend the static network model to the dynamic case
by combining the GLM with a state transition model and the EKF. This
is followed by describing the monitoring procedure for temporal change
detection. Next, performance of the proposed method is evaluated using
simulated network streams. A real\textendash world example based on
Enron corpus is then provided as a case study. Finally, the paper
is concluded in the last section.

\section*{Overview of the Proposed Methodology}

\begin{figure}
\centering{}\includegraphics[bb=50bp 60bp 860bp 500bp,clip,width=0.75\columnwidth]{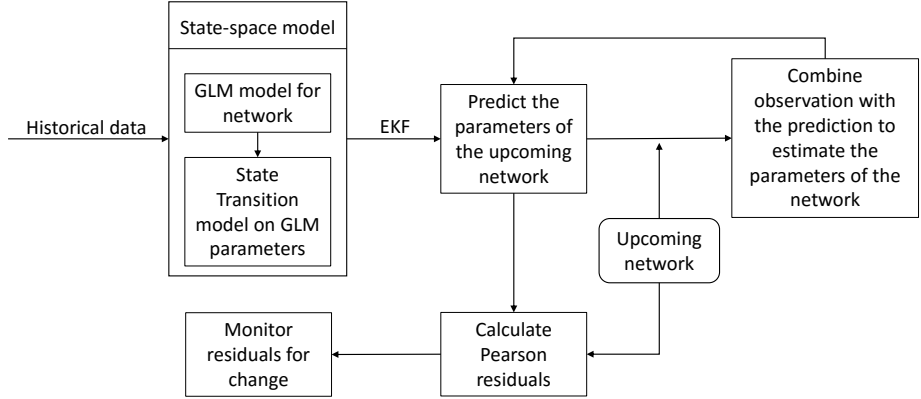}\caption{Overall procedure of network modeling and change detection\label{fig:Overall-procedure}}
\end{figure}

The overview of the proposed methodology for network modeling and
monitoring is given in Figure \ref{fig:Overall-procedure}. We begin
with modeling static attributed networks using GLM, in which the connectivity
of any pair of nodes is estimated by a function of the similarity
of their attributes. As GLM provides a general regression framework
for exponential family distributions, it can be used to model a broad
type of node connectivities. For example, for modeling the existence,
rate, and volume of communications among nodes, Bernoulli, Poisson,
and normal or gamma distributions can be used, respectively. Next,
to model the inherent dynamic of a network stream, we construct a
state\textendash space model (SSM) on the parameters of the GLM (i.e.
regression coefficients) and use the EKF to estimate and update model
parameters over time. In this procedure, we employ the estimation
of time $t-1$ to generate a one\textendash step\textendash ahead
prediction of the network at time $t$, which is used to compute Pearson
residuals \citep{myers2012generalized}. Finally, an EWMA chart is
constructed to monitor the Pearson residuals, and to detect abrupt
changes. 

In this paper, we use the following notations. We denote an observed
network at time $t$ with its corresponding adjacency matrix denoted
by $\boldsymbol{W}_{t}=\left[w_{ij,t}\right]$, where $w_{ij,t}$
is the weight of an edge connecting nodes $i$ and $j$ at the time
point $t$. In the case of binary networks, $w_{ij,t}=1$ if an edge
connects nodes $i$ and $j$, and $w_{ij,t}=0$ otherwise. We denote
the set of all networks up to time $t$ by $\boldsymbol{W}_{(t)}=\left\{ \boldsymbol{W}_{1},\boldsymbol{W}_{2},\cdots,\boldsymbol{W}_{t}\right\} $.
In general, we assume that networks are directed and there is no self-edge
in the networks, i.e. $w_{ii,t}=0$ for any $i$ and $t$. Here, without
loss of generality, we assume the total number of nodes remains fixed
over time, and denote the number of possible edges in a network by
$m=n\left(n-1\right)$, where $n$ is number of nodes. We suppose
that networks are attributed and use $\boldsymbol{x}_{ij,t}$ to denote
a $p$\textendash dimensional vector of attributes corresponding to
the edge between nodes $i$ and $j$ at time $t$. We let $\boldsymbol{X}_{t}=\left[\mathbf{1\;}\boldsymbol{x}_{ij,t}^{\mathtt{T}}\right]$
to denote the attribute matrix of size $m\times\left(p+1\right)$,
where $\mathbf{1}$ is an $m$\textendash dimensional vector of ones
corresponding to the intercept. Note that in the case of undirected
networks, $\boldsymbol{W}_{t}$ is a symmetric matrix and $\boldsymbol{x}_{ij,t}=\boldsymbol{x}_{ji,t}$.
We also denote the vectorized representation of the adjacency matrix
with $\boldsymbol{w}_{t}=vec\left(\left[w_{ij,t}\right]\right)$ for
$i\neq j$. Here, the operator $vec\left(.\right)$ transforms the
matrix $\boldsymbol{W}_{t}=\left[w_{ij,t}\right]$ into a vector of
size $m$. Finally, we denote the parameters of a network that statistically
relates $\boldsymbol{w}_{t}$ to $\boldsymbol{X}_{t}$ with a $p+1$
dimensional vector, $\boldsymbol{\beta}_{t}$.

\section*{Modeling of Attributed Network Streams}

We begin with modeling a given attributed network $\boldsymbol{W}_{t}$
at an arbitrary time $t$ using GLM. Detailed concepts and properties
of GLM can be found in \citep{nelder1972generalized}. Given a set
of attributes, we parametrize a network $\boldsymbol{W}_{t}$ by a
vector of parameters $\boldsymbol{\beta}_{t}$, which transforms the
network into its attribute space. This model assumes that every entry
of an adjacency matrix $\boldsymbol{W}_{t}$ is an independent realization
of an exponential family distribution (e.g. Bernoulli or Poisson)
with parameter $\theta_{ij,t}$ \textcolor{black}{given the attributes
of an edge}. This parameter denotes the probability of forming an
edge between nodes $i$ and $j$ in a given binary network, or represents
the average communication rate between two nodes in a given weighted
network, at time $t$. GLM models the parameters $\theta_{ij,t}$
as a function of a linear combination of attributes, i.e., $\boldsymbol{\theta}_{t}=vec\left(\left[\theta_{ij,t}\right]\right)=g\left(\boldsymbol{X}_{t}\boldsymbol{\beta}_{t}\right)$,
where $g:\mathbb{R\rightarrow\mathbb{R}}$ is an appropriate link
function that depends on the type of probability distribution. Examples
of link functions are $g\left(x\right)=\frac{1}{1+\exp\left(-x\right)}$
and $g\left(x\right)=e^{x}$ that can be used for modeling binary
and weighted networks, receptively. To find the maximum likelihood
estimate of the parameter $\boldsymbol{\beta}_{t}$ and its covariance
matrix denoted as $\boldsymbol{P}_{t}$, one can use iteratively reweighted
least square (IRWLS) algorithm \citep{nelder1972generalized} provided
in the Appendix.

The static attributed network model estimates the network parameter
$\boldsymbol{\beta}_{t}$, through GLM, for a given adjacency matrix
and a set of attributes at time $t$. However, this approach does
not consider the information of previously observed network snapshots,
and consequently cannot capture network dynamics in the network stream
$\boldsymbol{W}_{(t)}$. To address this issue, we integrate GLM with
SSM that incorporates the information of prior networks for a more
accurate estimation of model parameters. Specifically, we consider
$\boldsymbol{\beta}_{t}$ as the state of the system that generates
noisy observations through an observation equation given by 
\begin{equation}
\begin{cases}
w_{ij,t}\sim f\left(\theta_{ij,t}\right)\\
\boldsymbol{\theta}_{t}=vec\left(\left[\theta_{ij,t}\right]\right)=g\left(\boldsymbol{X}_{t}\boldsymbol{\beta}_{t}\right)
\end{cases},\label{eq:observation model}
\end{equation}
where $f\left(\theta_{ij,t}\right)$ denotes the probability density
function of an exponential family distribution with parameter $\theta_{ij,t}$.
The observation equation connects the parameters of the exponential
family distribution to the network attributes and facilitates transforming
the network to its attribute space. To complete the SSM, we consider
the following linear state transition equation: 
\begin{equation}
\boldsymbol{\beta}_{t}=\boldsymbol{F}\boldsymbol{\beta}_{t-1}+\boldsymbol{\xi}+\boldsymbol{\epsilon}_{t},\label{eq:state}
\end{equation}
where $\boldsymbol{F}$ is the state transition matrix, $\boldsymbol{\xi}$
is a vector of constants that determines the mean value of the parameter
$\boldsymbol{\beta}_{t}$, and $\boldsymbol{\epsilon}_{t}$ is the
white noise assumed to follow a Gaussian distribution with mean zero
and covariance matrix $\boldsymbol{Q}_{t}$. The transition matrix
$\boldsymbol{F}$ is assumed either to be known, or to be estimated
using system identification techniques \citep{ljung1998system}. To
estimate the model parameters $\boldsymbol{\beta}_{t}$, an inference
procedure is required. If the SSM were linear, the Kalman filter (KF)
procedure could achieve the optimal estimate of the states in terms
of the least square error \citep{kalman1960new}. However, clearly
the observation model in (\ref{eq:observation model}) is nonlinear.
Therefore, we employ the EKF shown to be effective in incorporating
nonlinearity in parameter estimation \citep{Fahrmeir1991}. Similar
to KF, EKF provides a recursive estimation procedure that only uses
the current network snapshot (at time \textit{$t$}) and the previous
parameter estimates (at time \textit{$t-1$}) to update the parameter
estimates at time $t$. This significantly reduces the computation
burden and avoids the out\textendash of\textendash memory issue as
the network stream grows large. In the following sub-sections, we
first describe the EKF approach, and then we present an approximate
but simpler alternative estimation procedure based on the KF framework
that replaces the nonlinear observation equation with a linear one.

\subsection*{Dynamic Estimation via Extended Kalman Filter }

Let $\boldsymbol{\beta}_{t|t-1}$ be the Kalman prediction of $\boldsymbol{\beta}_{t}$
given all the previous observations $\boldsymbol{W}_{\left(t-1\right)}$,
and let $\boldsymbol{\beta}_{t|t}$ be the Kalman estimation of $\boldsymbol{\beta}_{t}$
given all the observations up to time $t$, i.e., $\boldsymbol{W}_{\left(t\right)}$.
Similarly, let $\boldsymbol{P}_{t|t-1}$ and $\boldsymbol{P}_{t|t}$
denote Kalman prediction and estimation of the covariance matrix of
coefficients, $\boldsymbol{P}_{t}$. The EKF linearizes the observation
equation about the predicted state $\boldsymbol{\beta}_{t|t-1}$,
using the Taylor expansion to achieve a sub\textendash optimal estimate
of the state value. Let $\boldsymbol{G}_{t}$ denotes the Jacobian
of $g$ evaluated at $\boldsymbol{\beta}_{t|t-1}$. This matrix is
a $\left(p+1\right)\times m$ matrix obtained by $\boldsymbol{G}_{t}=\left.\frac{\partial g}{\partial\boldsymbol{\beta}_{t}}\right|_{\boldsymbol{\beta}_{t|t-1}}$.
Thus, the EKF prediction equations of the state $\boldsymbol{\beta}_{t}$
and its covariance matrix $\boldsymbol{P}_{t}$ for the SSM given
in (\ref{eq:observation model}) and (\ref{eq:state}) can be written
as follows \citep{Fahrmeir1991}: 
\begin{equation}
\boldsymbol{\beta}_{t|t-1}=\boldsymbol{F}\boldsymbol{\beta}_{t-1|t-1}+\boldsymbol{\xi},\label{eq:EKFstatePred}
\end{equation}
\begin{equation}
\boldsymbol{P}_{t|t-1}=\boldsymbol{F}\boldsymbol{P}_{t-1|t-1}\boldsymbol{F}^{\mathtt{T}}+\boldsymbol{Q}_{t}.\label{eq:EKF var pred}
\end{equation}
The prediction of the observation at time $t$, i.e., $\boldsymbol{w}_{t|t-1}$,
is given by the observation equation,

\begin{equation}
\boldsymbol{w}_{t|t-1}=g\left(\boldsymbol{X}_{t}\boldsymbol{\beta}_{t|t-1}\right),\label{eq:EKF obs equation}
\end{equation}
and consequently Kalman estimates are obtained by 
\begin{equation}
\boldsymbol{\beta}_{t|t}=\boldsymbol{\beta}_{t|t-1}+\boldsymbol{K}_{t}\left(\boldsymbol{w}_{t}-\boldsymbol{w}_{t|t-1}\right),\label{eq:EKF estiamte}
\end{equation}
\[
\boldsymbol{P}_{t|t}=\left(\boldsymbol{I}-\boldsymbol{K}_{t}\boldsymbol{G}_{t}\right)\boldsymbol{P}_{t|t-1},
\]
where $\boldsymbol{K}_{t}$ is the Kalman gain given by $\boldsymbol{K}_{t}=\boldsymbol{P}_{t|t-1}\boldsymbol{G}_{t}\left(\boldsymbol{G}_{t}^{\mathtt{T}}\boldsymbol{P}_{t|t-1}\boldsymbol{G}_{t}+\boldsymbol{R}_{t}\right)^{-1}$.
Here, $\boldsymbol{R}_{t}$ is an $m\times m$ diagonal matrix, which
represents the variance of observations and is estimated based on
the underlying network distribution and the observation prediction
$\boldsymbol{w}_{t|t-1}$. The estimated parameter $\boldsymbol{\beta}_{t|t}$
provides a sub\textendash optimal estimate of the network parameters
at time $t$ given a sequence of networks $\boldsymbol{W}_{\left(t\right)}$
\citep{Fahrmeir1991}. As examples, the detailed information and equations
of the estimation procedure for binary and weighted networks are given
in subsequent sections. 

\subsubsection*{Binary Networks}

Consider a binary network at an arbitrary time $t$, i.e., $\boldsymbol{W}_{t}$.
The binary network can be parametrized by a set of parameters $\theta_{ij,t}$
that denotes the probability of forming an edge between nodes $i$
and $j$. Thus, given two nodes, each element of the adjacency matrix
$\boldsymbol{W}_{t}$ is a realization of a Bernoulli distribution
with parameter $\theta_{ij,t}$, i.e., $w_{ij,t}\sim\textrm{Bernoulli}\left(\theta_{ij,t}\right)$.
Let the logistic function $g\left(x\right)=\frac{1}{1+\exp\left(-x\right)}$
be the appropriate link function related to the Bernoulli distribution.
Then, the ML estimate of the parameter $\boldsymbol{\beta}_{t}$ is
achieved by maximizing the following log-likelihood function using
the IRWLS given in the appendix: 

\[
l\left(\boldsymbol{\beta}_{t},\boldsymbol{X}_{t};\boldsymbol{w}_{t}\right)=\boldsymbol{w}_{t}^{\mathtt{T}}\log\left(\textrm{div}\left(\boldsymbol{1},\boldsymbol{1}+\exp\left(-\boldsymbol{X}_{t}\boldsymbol{\beta}_{t}\right)\right)\right)+\left(\boldsymbol{1}-\boldsymbol{w}_{t}\right)^{\mathtt{T}}\log\left(\textrm{div}\left(\exp\left(-\boldsymbol{X}_{t}\boldsymbol{\beta}_{t}\right),\boldsymbol{1}+\exp\left(-\boldsymbol{X}_{t}\boldsymbol{\beta}_{t}\right)\right)\right),
\]
where $\mathbf{1}$ is a vector of size $m$ whose elements are $1$,
$\textrm{div}\left(\boldsymbol{x},\boldsymbol{y}\right)$ is the element
by element division of vectors $\boldsymbol{x}$ and $\boldsymbol{y}$,
and the functions $\log\left(.\right)$ and $\exp\left(.\right)$
operate element\textendash wise on the input vectors. Now, assuming
the parameter $\boldsymbol{\beta}_{t}$ follows the state transition
equation given in (\ref{eq:state}), the Jacobian matrix $G_{t}$
is computed by 

\[
\boldsymbol{G}_{t}=\left.\frac{\partial}{\partial\boldsymbol{\beta}_{t}}\left(\textrm{div}\left(\boldsymbol{1},\boldsymbol{1}+\exp\left(-\boldsymbol{X}_{t}\boldsymbol{\beta}_{t}\right)\right)\right)\right|_{\boldsymbol{\beta}_{t|t-1}}=\boldsymbol{X}_{t}^{\mathtt{T}}\textrm{diag}\left(\exp\left(-\boldsymbol{X}_{t}\boldsymbol{\beta}_{t|t-1}\right)\right)\left(\textrm{diag}\left(\mathbf{1}+\exp\left(-\boldsymbol{X}_{t}\boldsymbol{\beta}_{t|t-1}\right)\right)\right)^{-2},
\]
where, $\textrm{diag}(\mathbf{x})$ is a diagonal matrix whose diagonal
elements are the elements of a vector $\mathbf{x}$. The estimation
of the network parameters $\beta_{t}$ is then achieved through the
equation (\ref{eq:EKF estiamte}). The covariance matrix of observations
$\boldsymbol{R}_{t}$ is an $m\times m$ diagonal matrix whose diagonal
element is obtained by $\left[r_{uu,t}\right]=w_{u,t|t-1}\left(1-w_{u,t|t-1}\right)$,
where $w_{u,t|t-1}$ refers to the $u^{th}$ element of the prediction
vector $\boldsymbol{w}_{t|t-1}$ computed in (\ref{eq:EKF obs equation}).

\subsubsection*{Weighted Networks}

Consider a weighted network at an arbitrary time $t$, i.e., $\boldsymbol{W}_{t}$,
in which weights represent the number of communications between two
nodes. This network is parametrized by a set of parameters $\theta_{ij,t}$,
denoting the communication rate between two nodes $i$ and $j$. Specifically,
we assume given two nodes, each element of the adjacency matrix $\boldsymbol{W}_{t}$
is a realization of a Poisson distribution with parameter $\theta_{ij,t}$,
i.e., $w_{ij,t}\sim\textrm{Poisson}\left(\theta_{ij,t}\right)$. Let
$g\left(x\right)=e^{x}$ be the link function related to the Poisson
distribution. Then, the ML estimate of the parameter $\boldsymbol{\beta}_{t}$
is achieved by maximizing the following log-likelihood function using
the IRWLS given in the appendix: 
\[
l\left(\boldsymbol{\beta}_{t},\boldsymbol{X}_{t};\boldsymbol{w}_{t}\right)=\boldsymbol{w}_{t}^{\mathtt{T}}\exp\left(-\boldsymbol{X}_{t}\boldsymbol{\beta}_{t}\right)-\mathbf{1^{\mathtt{T}}}\exp\left(-\boldsymbol{X}_{t}\boldsymbol{\beta}_{t}\right),
\]
where $\mathbf{1}$ is a vector of size $m$ whose elements are 1.
Now, suppose that the parameter $\boldsymbol{\beta}_{t}$ follows
a linear state transition equation (\ref{eq:state}), then $\boldsymbol{G}_{t}$
is given by 
\begin{align*}
\boldsymbol{G}_{t} & =\left.\frac{\partial}{\partial\boldsymbol{\beta}_{t}}\left(\exp\left(-\boldsymbol{X}_{t}\boldsymbol{\beta}_{t}\right)\right)\right|_{\boldsymbol{\beta}_{t|t-1}}=\boldsymbol{X}_{t}^{\mathtt{T}}\textrm{diag}\left(\exp\left(\boldsymbol{X}_{t}\boldsymbol{\beta}_{t|t-1}\right)\right).
\end{align*}
Given $\boldsymbol{G}_{t}$, the estimation of the network parameters
$\boldsymbol{\beta}_{t}$ is achieved through equation (\ref{eq:EKF estiamte}).
The covariance matrix of observations $\boldsymbol{R}_{t}$ is an
$m\times m$ diagonal matrix with diagonal elements $\left[r_{uu,t}\right]=w_{u,t|t-1}$,
where $w_{u,t|t-1}$ refers to the $u^{th}$ element of the prediction
vector $w_{t|t-1}$ computed in (\ref{eq:EKF obs equation}).

\subsection*{Approximate Estimation via Kalman Filter\label{subsec:Approximate-Estimation}}

When the size of a network (adjacency matrix) is large, one may consider
replacing the nonlinear observation equation with a linear equation
with respect to $\boldsymbol{\beta}_{t}$. This linearization can
significantly simplify the inference procedure. Let denote the static
ML estimate of a network parameter at time\textit{ $t$} by $\hat{\boldsymbol{\beta}_{t}}$.
It is known that $\hat{\boldsymbol{\beta}_{t}}$ asymptotically follows
a Gaussian distribution with mean $\boldsymbol{\beta}_{t}$ and the
covariance matrix $\boldsymbol{P}_{t}$ \citep{myers2012generalized}.
Therefore, as the link function for the Gaussian distribution is linear,
one can rewrite the observation equation in a linear form as $\hat{\boldsymbol{\beta}_{t}}=\boldsymbol{\beta}_{t}+\boldsymbol{v}_{t}$
where $\boldsymbol{v}_{t}$ is a random $\left(p+1\right)$\textendash vector
of a zero\textendash mean Gaussian distribution with the covariance
matrix of $\boldsymbol{R}_{t}$ . Consequently, the SSM becomes linear
as follows: 

\begin{equation}
\begin{cases}
\boldsymbol{\beta}_{t}=\boldsymbol{F}\boldsymbol{\beta}_{t-1}+\boldsymbol{\xi}+\boldsymbol{\epsilon}_{t},\\
\hat{\boldsymbol{\beta}_{t}}=\boldsymbol{\beta}_{t}+\boldsymbol{v}_{t}.
\end{cases}\label{eq:state-1}
\end{equation}

With the linear observation model, KF can be used to predict and estimate
the parameters $\boldsymbol{\beta}_{t}$ by the following recursive
equations:

\begin{equation}
\boldsymbol{\beta}_{t|t}=\boldsymbol{\beta}_{t|t-1}+\boldsymbol{K}_{t}\left(\hat{\boldsymbol{\beta}_{t}}-\boldsymbol{\beta}_{t|t-1}\right),\label{eq:linearKalman1}
\end{equation}

\begin{equation}
\boldsymbol{K}_{t}=\boldsymbol{P}_{t|t-1}\left(\boldsymbol{P}_{t|t-1}+\boldsymbol{R}_{t}\right)^{-1},\label{eq:kalmanlinear2}
\end{equation}
\begin{equation}
\boldsymbol{P}_{t|t}=\left(\boldsymbol{I}-\boldsymbol{K}_{t}\right)\boldsymbol{P}_{t|t-1},\label{eq:KalmanLinear3}
\end{equation}
where $\boldsymbol{\beta}_{t|t-1}$ and $\boldsymbol{P}_{t|t-1}$
are obtained by (\ref{eq:EKFstatePred}) and (\ref{eq:EKF var pred}).

\section*{Monitoring of Dynamic Network Streams}

\textcolor{black}{Our proposed modeling approach provides a means
for estimating and updating the parameters of a dynamic attributed
network stream. That is, it can capture the network dynamics and the
autocorrelation structure among the network snapshots}. This section
proposes a monitoring procedure for detecting abrupt changes in a
stream of attributed networks caused by a shock different from the
underlying dynamic mechanism of the network stream. In this paper,
we focus on Phase II monitoring where we assume that the in-control
model and its initial parameters $\boldsymbol{F}$, $\boldsymbol{\xi}$,
etc., are either known or can be estimated from an in-control sequence
of attributed networks denoted by $\boldsymbol{W}_{\left(T\right)}=\left\{ \boldsymbol{W}_{1},\boldsymbol{W}_{2},\cdots,\boldsymbol{W}_{T}\right\} $.
For each incoming network snapshot, the network parameters are predicted
using the EKF or KF equations discussed before. Then, the vector of
updated parameters $\boldsymbol{\beta}_{t+1|t}$ is used to predict
the adjacency matrix at time $t$ for $t=T+1,...$, using the appropriate
link function, i.e., $\boldsymbol{w}_{t+1|t}=g\left(\boldsymbol{X}_{t}\boldsymbol{\beta}_{t+1|t}\right)$\textcolor{black}{,
where $\boldsymbol{w}_{t+1|t}$ is the vectorized version of the corresponding
adjacency matrix $W_{t+1|t}$ and is of size $m$.} Next, the vector
of Pearson residuals of size $m$ denoted by $\boldsymbol{r}_{t+1}$
is computed by \citep{myers2012generalized}
\[
r_{i,t+1}=\frac{w_{i,t+1}-w_{i,t+1|t}}{\sqrt{var\left(w_{i,t+1|t}\right)}};\:\mathrm{for}\:t=T,T+1,...,
\]
\textcolor{black}{where, $w_{i,t+1}$ is the the $i^{th}$ element
of vectorized version of the adjacency matrix observed at time $t+1$,
$w_{i,t+1|t}$ is the $i^{th}$ element of vectorized version of predicted
adjacency matrix, and $r_{i,t+1}$} is the $i^{th}$ element of the
residual vector $\boldsymbol{r}_{t+1}$. The prediction variance $var\left(w_{i,t+1|t}\right)$
depends on the distribution of the edge formation in the network.
For example, in the case of binary network the variance is given by
$var\left(w_{i,t+1|t}\right)=w_{i,t+1|t}\left(1-w_{i,t+1|t}\right)$,
and in the case of weighted networks, it is given by $var\left(w_{i,t+1|t}\right)=w_{i,t+1|t}$. 

If the process is in-control, Pearson residuals, $r_{i,t+1}$, asymptotically\textcolor{blue}{{}
}\textcolor{black}{and independently} follow a standard normal distribution.
Therefore, one can construct an EWMA control chart to test the hypothesis
whether $E(r_{i,t+1})=0$ fo\textcolor{black}{r $t=T,T+1,...$ and
$i=1,2,\cdots,m$. However, instead of monitoring each $r_{i,t+1}$,
we monitor $\bar{r}_{t+1}=\frac{1}{m}{\textstyle \sum_{i=1}^{m}}r_{i,t+1}$
that also follows a normal distribution with mean zero, in the case
of in-control data.} The EWMA statistic at time $t+1$ is computed
by $z_{t+1}=\lambda\bar{r}_{t+1}+\left(1-\lambda\right)z_{t}$, where
$\lambda$ is the weight factor, and $z_{0}=0.$ The upper control
limit (UCL) and the lower control limit (LCL) of the EWMA control
chart at time $t$ are given by 
\[
UCL_{t}=l\times s\sqrt{\frac{\lambda}{1-\lambda}\left(1-\left(1-\lambda\right)^{2t}\right)},
\]
\[
LCL_{t}=-l\times s\sqrt{\frac{\lambda}{1-\lambda}\left(1-\left(1-\lambda\right)^{2t}\right)},
\]
where $s$ is the in\textendash control standard deviation of $\bar{r}_{t}$
and should be estimated based on the reference set $\boldsymbol{W}_{\left(T\right)}$,
and $l$ and $\lambda$ are obtained through Monte Carlo simulations
so that a desired in\textendash control average run length (ARL) is
achieved as detailed in the next section. If $z_{T+1}\geq UCL_{T+1}$
or $z_{T+1}\leq LCL_{T+1}$, we reject the null hypothesis, indicating
a change has occurred in the network stream. 

\section*{Performance Evaluation Using Simulation}

This section evaluates the performance of the proposed methodology
in detecting abrupt changes by using simulated streams of networks.
We compare the performance of the proposed method with two benchmark
approaches. The first benchmark is the static GLM that at each sampling
time fits a GLM using only the current network data. In this setting,
we consider the current network estimate as a prediction for the next
network snapshot to calculate residuals. The second benchmark is a
variant method for handling network dynamics proposed by \citet{azarnoush2016monitoring}.
They suggested to aggregate the last $l_{w}$ observed networks in
order to predict the upcoming network. The predicted values from each
method are then used for calculating the Pearson residuals. The out\textendash of\textendash control
average and standard deviation of run lengths are considered as performance
measures for comparison. 

For this simulation study, we consider a stream of communication networks
of college students. Each network snapshot contains the communication
information of $50$ students during a week. The age of students ranging
from $20$ to $40$ is one of the two attributes used in this simulation,
and is denoted by $a$. We assume that in an in-control situation
the number of contacts between two students $i$ and $j$ depends
on their age difference $a_{ij}=|a_{i}-a_{j}|$. That is, two students
with smaller age difference are more likely to contact each other.
Furthermore, we suppose five students are members of an association
that holds cultural events and requests the active members to promote
the events. Therefore, when the association members are active, they
have an excessive communication with their network members. The association
membership is modeled by a binary variable $c_{ij}$, and is considered
as the second attribute. This attribute is one if its corresponding
edge is adjacent to an association member, and is zero otherwise.
Therefore, the vector of edge attributes denoted by $\boldsymbol{x}_{ij}=[a_{ij};c_{ij}]$,
consists of a continuous and a binary variable. 

We generate two types of networks: First, we consider binary networks
in which an edge between two nodes at week $t$ represents at least
one contact between two students during that week. Second, we simulate
weighted networks, in which the weight of an edge encodes the number
of times two students contacted during a particular week. For simulating
a binary graph, we assume the probability of an edge between two nodes
is given by the following logistic function $\theta_{ij,t}=\frac{1}{1+\exp\left(-\boldsymbol{\beta}_{t}^{\mathtt{T}}\left[1;\boldsymbol{x}_{ij}\right]\right)}.$
In the case of weighed networks, we suppose that weights follow a
Poisson distribution, in which the average number of contacts is given
by an exponential function as follows: $\theta_{ij,t}=\exp\left(\boldsymbol{\beta}_{t}^{\mathtt{T}}\left[1;\boldsymbol{x}_{ij}\right]\right)$.
Here, $\boldsymbol{\beta}_{t}^{\mathtt{T}}=\left[\beta_{1,t}\,\,\beta_{2,t}\,\,\beta_{3,t}\right]$
is the vector of the model parameters, where $\beta_{1,t}$ is an
intercept, and $\beta_{2,t}$ and $\beta_{3,t}$ represent the effect
of $a_{ij,t}$ and $c_{ij,t}$ on the probability (or weight), respectively.
The underlying dynamic of the communication network stream is generated
through a state transition model given by $\boldsymbol{\beta}_{t}=\boldsymbol{F}\boldsymbol{\beta}_{t-1}+\boldsymbol{\xi}+\boldsymbol{\epsilon}_{t}$.
In the simulations, we use $\boldsymbol{\beta}_{0}^{\mathtt{T}}=\left[-1\,\,0.05\,\,0\right]$
as initial state values, and $\boldsymbol{F}=0.7\boldsymbol{I}_{3\times3}$,
$\boldsymbol{\xi}^{\mathtt{T}}=\left[-0.6\,\,0.03\,\,0\right]$ as
the parameters of the state transition model. Furthermore, we assume
$\boldsymbol{\epsilon}_{t}\sim NID\left(0,\boldsymbol{Q}\right)$,
where $\boldsymbol{Q}$ is a $3\times3$ diagonal matrix with $Q_{11}=0.01$,
$Q_{22}=0.0001$, $Q_{33}=0.05$. We set the initial value of $\beta_{3,t}$
to zero, indicating that the association members are inactive at the
time. To generate a sequence of networks, we first generate a sequence
of state values $\left\{ \boldsymbol{\beta}_{t}\right\} _{t=1}^{100}$
through the state transition equation with initial values of $\boldsymbol{\beta}_{0}$.
Then, a sequence of edge formation rates $\left\{ \theta_{ij,t}\right\} $
is calculated through the corresponding link functions. When simulating
a binary network, we generate a random number from a Bernoulli distribution
with the calculated parameters $\theta_{ij,t}$ and place an edge
between two nodes if the generated value is equal to one. In the case
of weighted networks, we produce weights through a Poisson distribution
with mean $\theta_{ij,t}$. The generated sequence of networks is
considered as an input to the proposed methodology. 

To evaluate the performance of the proposed methodology for each network
type, we consider two scenarios where a change is imposed at time
$t=\tau$ by increasing or decreasing the value of $\beta_{i,\tau}$
by $\delta\sigma_{i}$, where $\delta$ is a constant representing
the magnitude of the change and $\sigma_{i}$ denotes the standard
deviation of the parameter $\beta_{i}$ prior to the change given
by $\sigma_{i}=\sqrt{\frac{Q_{ii}}{\left(1-F_{ii}\right)^{2}}}$.
The two scenarios are as follows:

\textbf{Scenario 1: }We assume at time $\tau=50$, students are graduating
and therefore the overall communication level is reduced. We model
this change by reducing the second parameter of the transition equation
by $\delta\sigma_{2}$. That is $\beta_{2,\tau}=F_{22}\beta_{2,\tau-1}+\xi_{2}+\epsilon_{2,\tau}-\delta\sigma_{2}$.
This scenario represents a global change affecting all nodes. 

\textbf{Scenario 2: }We suppose at time $\tau=50$, the student members
of the association decide to become active. We model this change by
increasing $\beta_{3}$ by $\delta\sigma_{3}$, i.e., $\beta_{3,\tau}=F_{33}\beta_{3,\tau-1}+\xi_{3}+\epsilon_{3,\tau}+\delta\sigma_{3}$.
This scenario represents a local change, in which only few of the
nodes have excessive communications with the entire network.

\begin{figure}
\centering{}%
\begin{tabular}{|c|c|}
\hline 
\multicolumn{2}{|c|}{Binary network}\tabularnewline
\hline 
Scenario 1 & Scenario 2\tabularnewline
\hline 
\subfloat[]{\includegraphics[width=0.45\columnwidth]{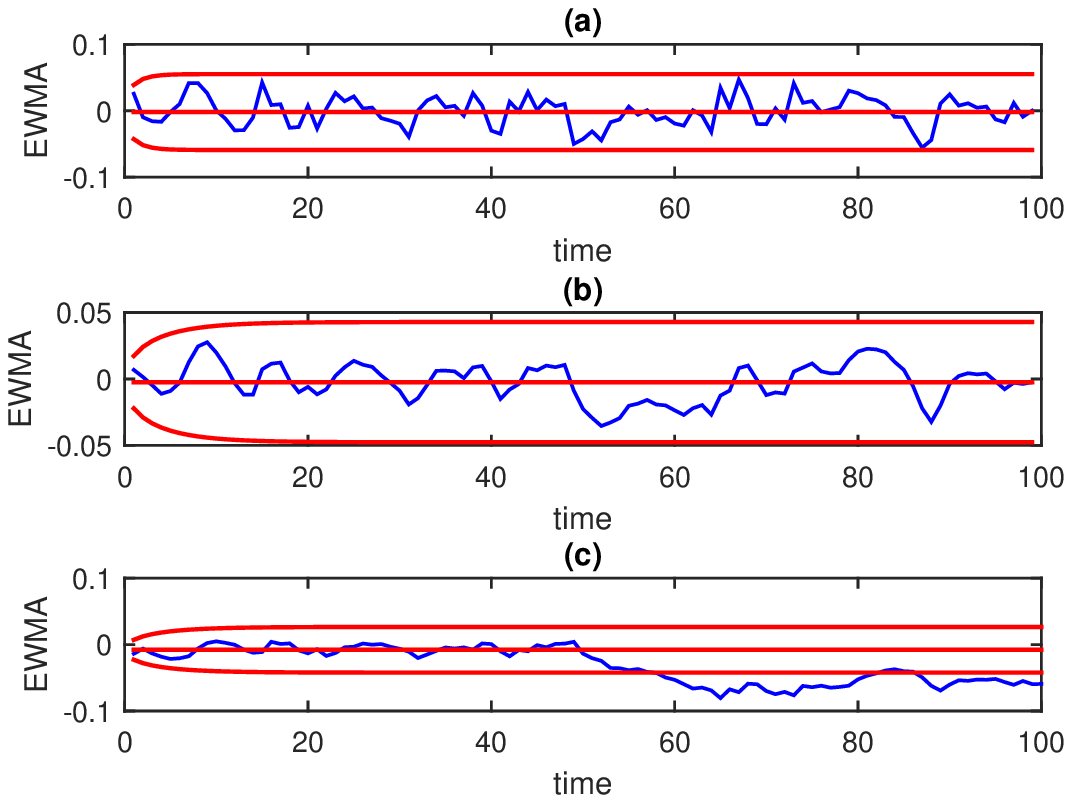}} & \subfloat[]{\includegraphics[width=0.45\columnwidth]{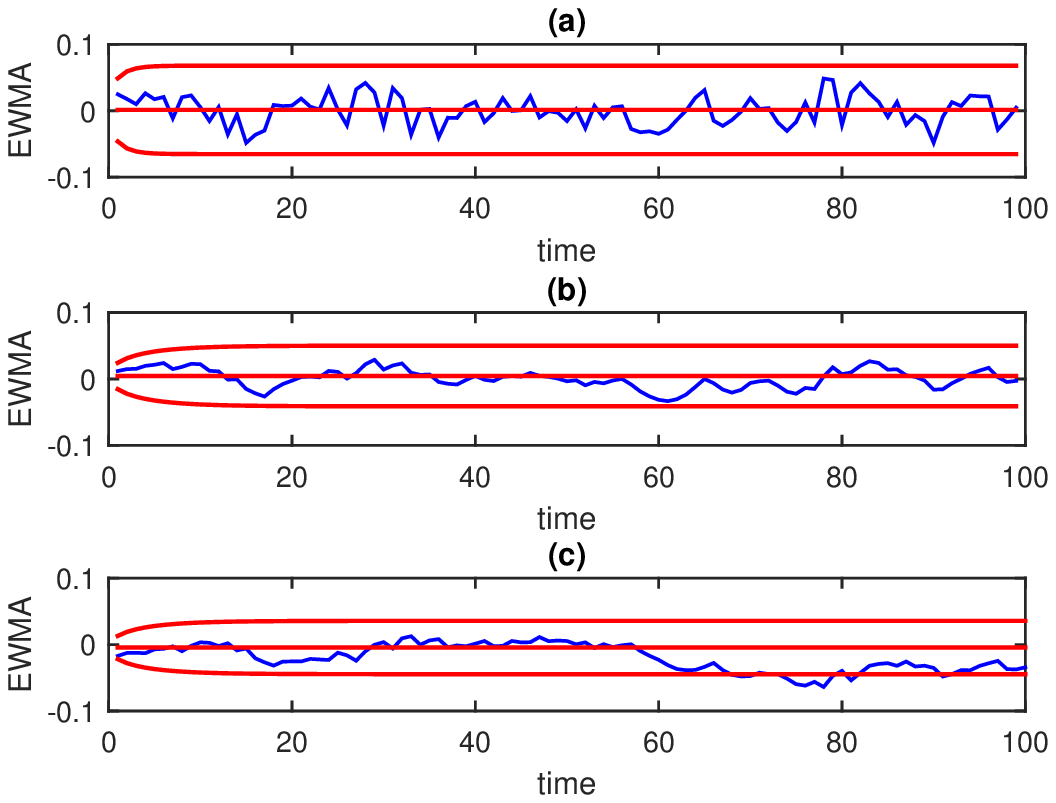}}\tabularnewline
\hline 
\multicolumn{2}{|c|}{Weighted network}\tabularnewline
\hline 
Scenario 1 & Scenario 2\tabularnewline
\hline 
\subfloat[]{\includegraphics[width=0.45\columnwidth]{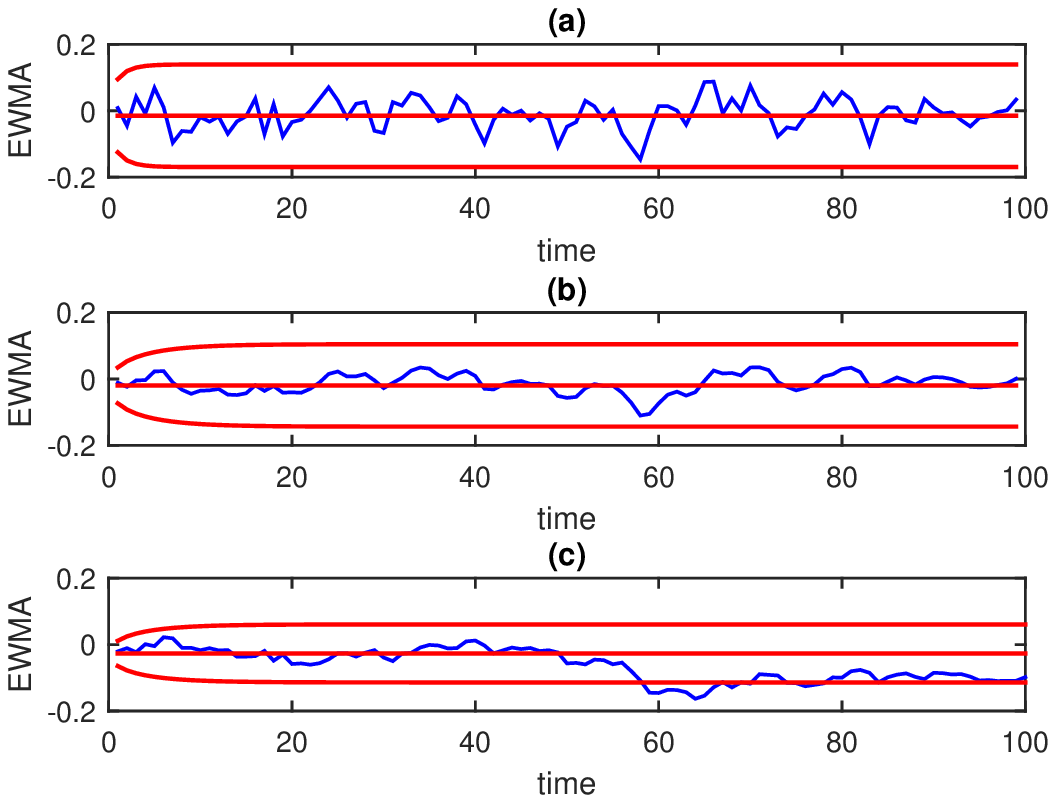}} & \subfloat[]{\includegraphics[width=0.45\columnwidth]{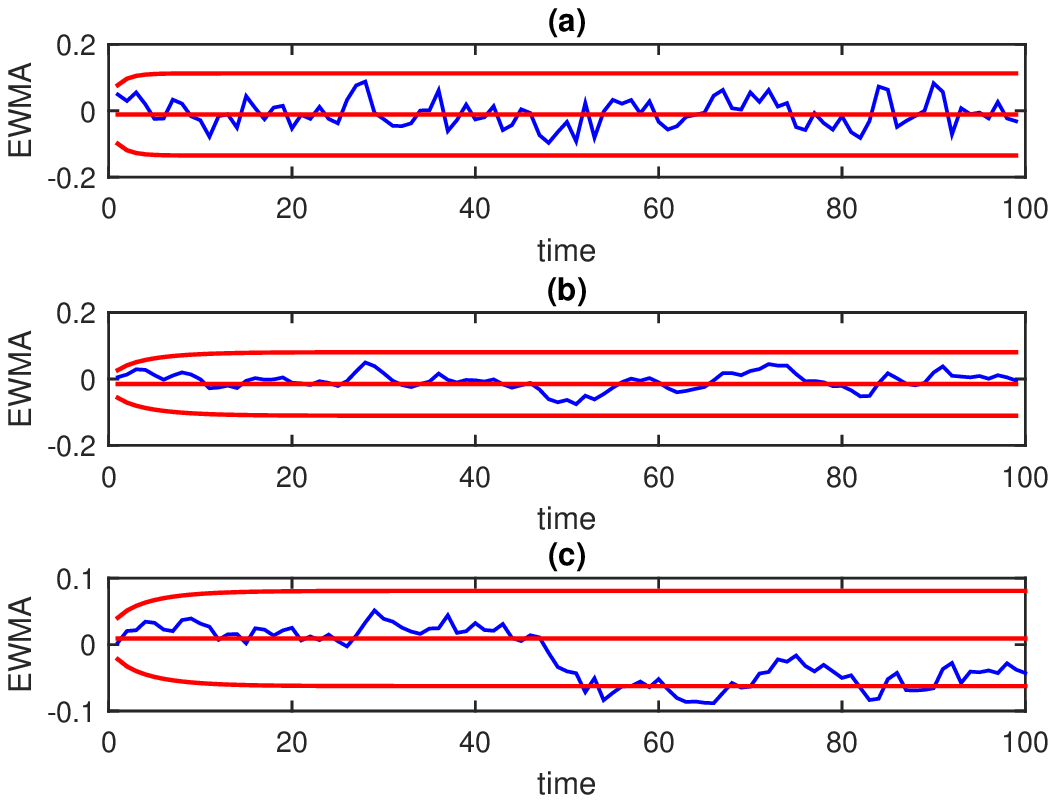}}\tabularnewline
\hline 
\end{tabular}\caption{Plot of EWMA statistics versus time. Shift of magnitude $\delta=2$
imposed at $\tau=50$. In each figure, (a), (b), and (c) represents
static model, sliding static model with $l_{w}=5$, and the dynamic
model. \label{fig:Plot-of-EWMA-allcases}}
\end{figure}

The control limits are calculated using the Monte\textendash Carlo
simulations to achieve the in\textendash control ARL ($ARL_{0}$ )
of $200$. These values are obtained through $2000$ replications
of the simulation and the in\textendash control sequence of $5000$
networks. First, a network sequence of length 5000 is generated based
on the in\textendash control model. This stream is used as a reference
set to estimate the standard deviation of the Pearson residuals, $s$.
Then, we generate 2000 in\textendash control network sequences of
length 5000 and determine the ARL for a given pair of $l$ and $\lambda$.
We repeat this procedure for several pairs of $l$ and $\lambda$
to find those resulting in $ARL_{0}$ of $200$. The resulting $l$
and $\lambda$ values for each method are reported in Table \ref{tab:Values-of-LandLambda}.
We use these parameters to identify the control limits for change
detection. For illustration, we plot the EWMA charts shown in the
Figure \ref{fig:Plot-of-EWMA-allcases}. This figure gives an example
of detection performance of each method in different scenarios. The
charts illustrate the EWMA values calculated based on the Pearson
residuals.
\begin{table}
\centering{}\caption{Values of $l$ and $\lambda$ that achieves $ARL_{0}=200$\label{tab:Values-of-LandLambda}}
\begin{tabular}{|c|c|c|c|}
\hline 
$\left(l,\lambda\right)$ & Static model & Sliding static model & Dynamic model\tabularnewline
\hline 
\hline 
Binary & $\left(1.79,0.3\right)$ & $\left(3.33,0.1\right)$ & $\left(2.44,0.1\right)$\tabularnewline
\hline 
Weighted & $\left(2.00,0.3\right)$ & $\left(3.61,0.1\right)$ & $\left(2.53,0.1\right)$\tabularnewline
\hline 
\end{tabular}
\end{table}

\begin{figure}[H]
\begin{centering}
\begin{tabular}{|c|c|}
\hline 
\multicolumn{2}{|c|}{{\footnotesize{}Binary network}}\tabularnewline
\hline 
{\footnotesize{}Scenario 1 (a)} & {\footnotesize{}Scenario 2 (b)}\tabularnewline
\hline 
\includegraphics[width=0.45\columnwidth]{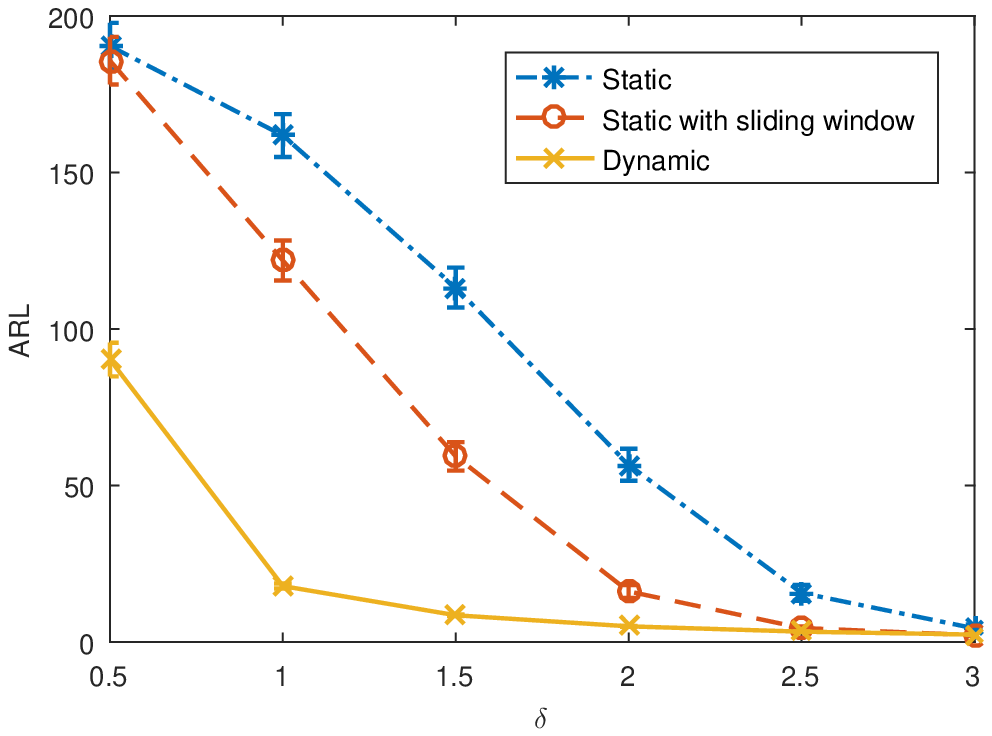} & \includegraphics[width=0.45\columnwidth]{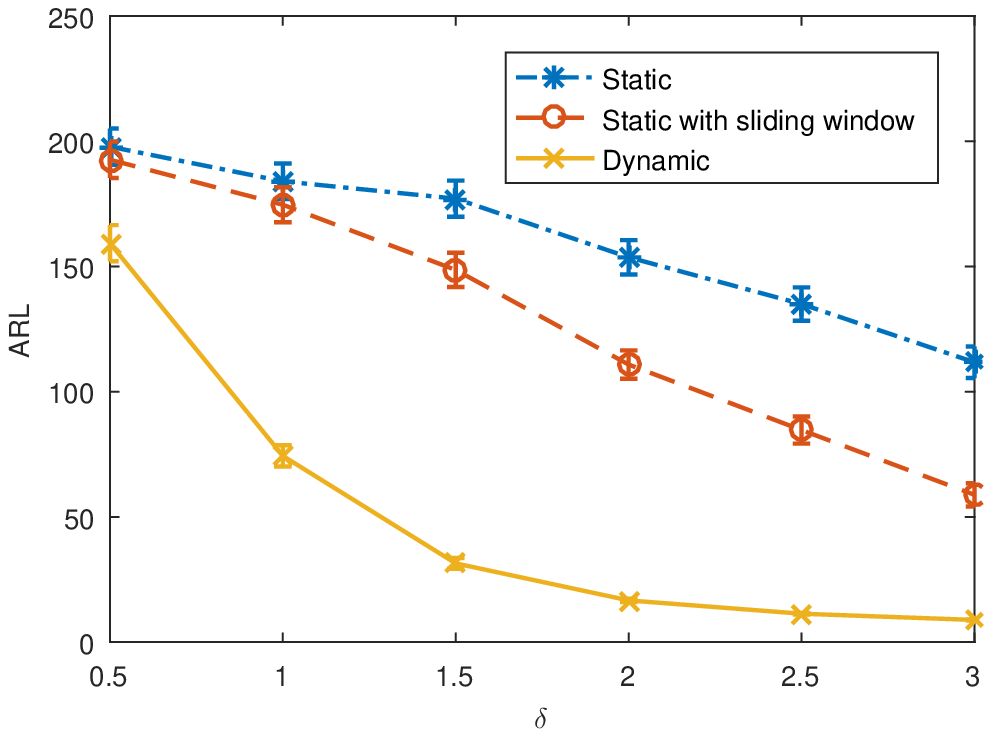}\tabularnewline
\hline 
\multicolumn{2}{|c|}{{\footnotesize{}Weighted network}}\tabularnewline
\hline 
{\footnotesize{}Scenario 1 (c)} & {\footnotesize{}Scenario 2 (d)}\tabularnewline
\hline 
\includegraphics[width=0.45\columnwidth]{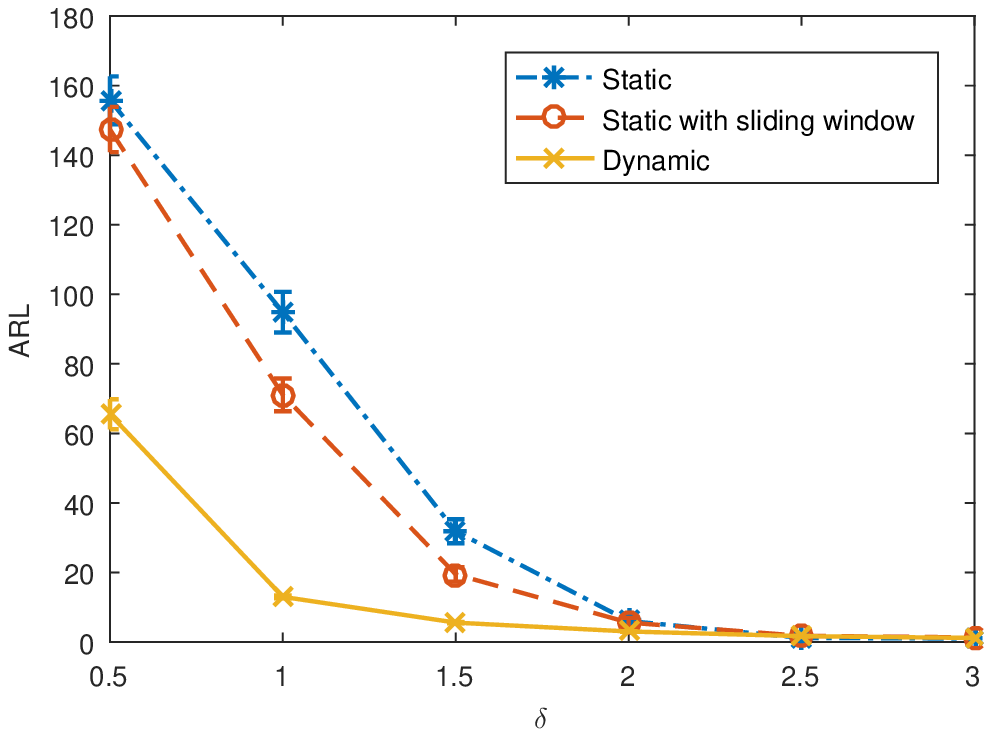} & \includegraphics[width=0.45\columnwidth]{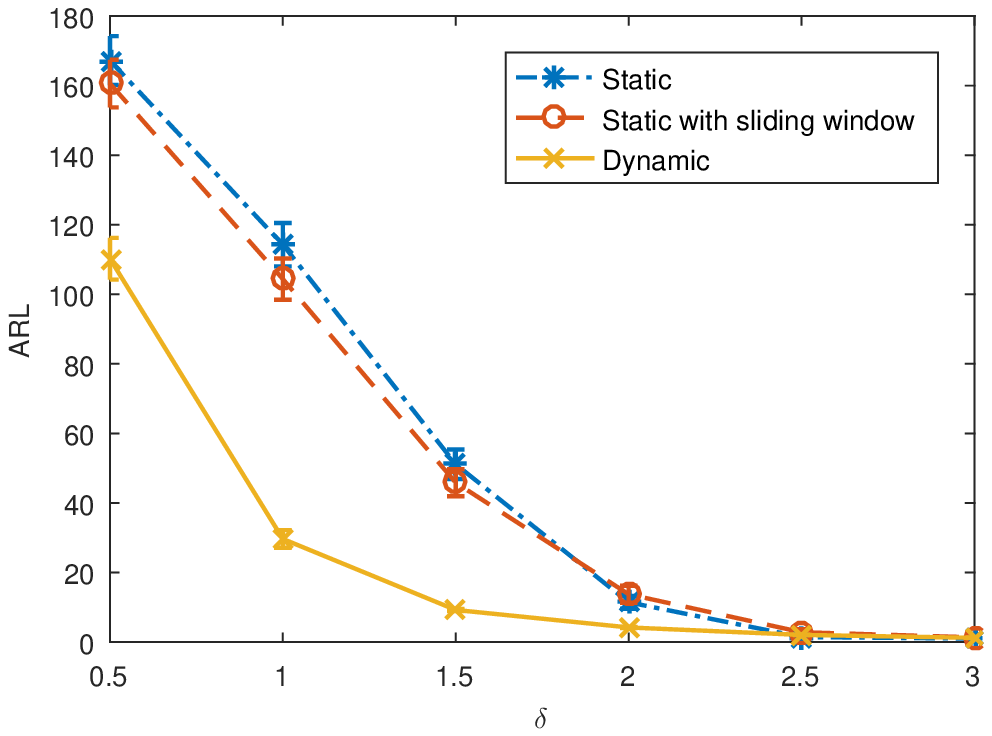}\tabularnewline
\hline 
\end{tabular}
\par\end{centering}
\caption{Average run length as a function of change magnitude for the proposed
modeling framework and the benchmarks\label{fig:Average-run-length-plots}}
\end{figure}

To compare the performance of the proposed method with the benchmarks
that ignore the dynamic evolution of networks, we compute the out\textendash of\textendash control
ARL ($ARL_{1}$) for each scenario using 2000 simulation replications.
Figure \ref{fig:Average-run-length-plots} along with Table \ref{tab:Out--of--control-average-run}
report the performance of the proposed and benchmark methods for each
network type and scenario. As it can be seen from the Figure, our
proposed method that considers the dynamic evolution of the network
can detect smaller changes (i.e. $\delta<2.5$) faster (with smaller
standard error) than both benchmark methods in all the cases. For
example, in Scenario 1, the $ARL_{1}\,\left(SERL\right)$ values of
our dynamic method for detecting a change with the magnitude of $\delta=1.5$
in binary network streams is 8.57 (0.144), while these values for
the static and sliding static methods are, respectively, 113.27 (3.869)
and 59.32 (2.770). This indicates that the dynamic method can detect
such a change around 15 and 9 times faster than the benchmark methods. 

The main reason that benchmark methods fail to detect small changes
is that these methods do not properly capture\textcolor{black}{{} autocorrelation
structure among the network snapshots }(i.e. gradual changes) caused
by network dynamics, resulting in a small $ARL_{0}$ (or large false
alarm rate). Therefore, by keeping $ARL_{0}$ close to 200, these
methods lose the detection power leading to large $ARL_{1}$. Another
reason for poor performance of the static method is that they have
a very short window of opportunity for detecting a change. In the
static method, we predict the next network only based on the current
network estimations. If a change appears upon the arrival of the next
network (i.e. new regime starts) the difference between the arriving
network and its prediction is larger than the previous cases, representing
a change. However, if this change is not detected at the change point,
the prediction of the future networks are made based on the estimation
of the networks that are already generated based on the new regime
and therefore the residuals become small. This short window of opportunity
in detection of changes is also reflected in the standard errors of
the benchmark methods, reported in Table \ref{tab:Out--of--control-average-run}.
The large standard error values indicate that a change is either detected
very quickly or never been detected by the benchmarks. For larger
changes (i.e. $\delta\geq2.5$) the performance of all three methods
are comparable, though the proposed method still has slightly smaller
detection delay in the case of binary networks. 

For the local change in binary networks (Scenario 2), the excessive
communication cannot be completely captured, due to the limited data.
This limitation is reflected in the detection power of all three methods.
That is, all three methods have larger $ARL_{1}$ values compared
with the corresponding global changes. Nevertheless, the dynamic method
significantly outperforms the benchmarks in detecting a local change
in binary networks. This again shows the importance of capturing the
network dynamics and its effect in change detection. When we model
the communication considering the number of contacts (weighted networks),
the local change is more apparent and can be detected by all three
methods when it is large enough. However, our proposed method can
still detect smaller local changes faster in weighted network streams.
\begin{table}[H]
\caption{Standard error ($\frac{\sigma_{ARL_{1}}}{\sqrt{n}}$ for $n=2000$)
as a function of change magnitude for the proposed modeling framework
and the benchmarks\label{tab:Out--of--control-average-run}}
\centering{}%
\begin{tabular}{|c|c|}
\hline 
\multicolumn{2}{|c|}{{\footnotesize{}Binary network}}\tabularnewline
\hline 
{\footnotesize{}Scenario 1} & {\footnotesize{}Scenario 2}\tabularnewline
\hline 
{\footnotesize{}}%
\begin{tabular}{|c|c|c|c|}
\hline 
{\footnotesize{}$\delta$} & {\footnotesize{}Static} & {\footnotesize{}Sliding static } & {\footnotesize{}Dynamic }\tabularnewline
\hline 
\hline 
{\footnotesize{}$0.5$} & {\footnotesize{}4.494} & {\footnotesize{}4.617} & {\footnotesize{}3.270}\tabularnewline
\hline 
{\footnotesize{}$1$} & {\footnotesize{}4.156} & {\footnotesize{}3.874} & {\footnotesize{}0.486}\tabularnewline
\hline 
{\footnotesize{}$1.5$} & {\footnotesize{}3.869} & {\footnotesize{}2.770} & {\footnotesize{}0.144}\tabularnewline
\hline 
{\footnotesize{}$2$} & {\footnotesize{}3.096} & {\footnotesize{}1.260} & {\footnotesize{}0.083}\tabularnewline
\hline 
{\footnotesize{}$2.5$} & {\footnotesize{}1.520} & {\footnotesize{}0.319} & {\footnotesize{}0.051}\tabularnewline
\hline 
{\footnotesize{}$3$} & {\footnotesize{}0.785} & {\footnotesize{}0.053} & {\footnotesize{}0.036}\tabularnewline
\hline 
\end{tabular} & {\footnotesize{}}%
\begin{tabular}{|c|c|c|c|}
\hline 
{\footnotesize{}$\delta$} & {\footnotesize{}Static } & {\footnotesize{}Sliding static } & {\footnotesize{}Dynamic }\tabularnewline
\hline 
\hline 
{\footnotesize{}$0.5$} & {\footnotesize{}4.406} & {\footnotesize{}4.416} & {\footnotesize{}4.387}\tabularnewline
\hline 
{\footnotesize{}$1$} & {\footnotesize{}4.319} & {\footnotesize{}4.256} & {\footnotesize{}2.580}\tabularnewline
\hline 
{\footnotesize{}$1.5$} & {\footnotesize{}4.391} & {\footnotesize{}4.161} & {\footnotesize{}1.298}\tabularnewline
\hline 
{\footnotesize{}$2$} & {\footnotesize{}4.171} & {\footnotesize{}3.444} & {\footnotesize{}0.403}\tabularnewline
\hline 
{\footnotesize{}$2.5$} & {\footnotesize{}4.053} & {\footnotesize{}3.298} & {\footnotesize{}0.189}\tabularnewline
\hline 
{\footnotesize{}$3$} & {\footnotesize{}3.828} & {\footnotesize{}2.821} & {\footnotesize{}0.145}\tabularnewline
\hline 
\end{tabular}\tabularnewline
\hline 
\multicolumn{2}{|c|}{{\footnotesize{}Weighted network}}\tabularnewline
\hline 
{\footnotesize{}Scenario 1} & {\footnotesize{}Scenario 2}\tabularnewline
\hline 
{\footnotesize{}}%
\begin{tabular}{|c|c|c|c|}
\hline 
{\footnotesize{}$\delta$} & {\footnotesize{}Static model} & {\footnotesize{}Sliding static } & {\footnotesize{}Dynamic }\tabularnewline
\hline 
\hline 
{\footnotesize{}$0.5$} & {\footnotesize{}4.184} & {\footnotesize{}3.954} & {\footnotesize{}2.633}\tabularnewline
\hline 
{\footnotesize{}$1$} & {\footnotesize{}3.543} & {\footnotesize{}2.860} & {\footnotesize{}0.257}\tabularnewline
\hline 
{\footnotesize{}$1.5$} & {\footnotesize{}2.110} & {\footnotesize{}1.262} & {\footnotesize{}0.109}\tabularnewline
\hline 
{\footnotesize{}$2$} & {\footnotesize{}0.735} & {\footnotesize{}0.828} & {\footnotesize{}0.057}\tabularnewline
\hline 
{\footnotesize{}$2.5$} & {\footnotesize{}0.056} & {\footnotesize{}0.027} & {\footnotesize{}0.031}\tabularnewline
\hline 
{\footnotesize{}$3$} & {\footnotesize{}0.020} & {\footnotesize{}0.019} & {\footnotesize{}0.014}\tabularnewline
\hline 
\end{tabular} & {\footnotesize{}}%
\begin{tabular}{|c|c|c|c|}
\hline 
{\footnotesize{}$\delta$} & {\footnotesize{}Static } & {\footnotesize{}Sliding static } & {\footnotesize{}Dynamic }\tabularnewline
\hline 
\hline 
{\footnotesize{}$0.5$} & {\footnotesize{}4.273} & {\footnotesize{}4.154} & {\footnotesize{}3.653}\tabularnewline
\hline 
{\footnotesize{}$1$} & {\footnotesize{}3.878} & {\footnotesize{}3.613} & {\footnotesize{}1.544}\tabularnewline
\hline 
{\footnotesize{}$1.5$} & {\footnotesize{}2.577} & {\footnotesize{}2.366} & {\footnotesize{}0.178}\tabularnewline
\hline 
{\footnotesize{}$2$} & {\footnotesize{}1.122} & {\footnotesize{}1.496} & {\footnotesize{}0.090}\tabularnewline
\hline 
{\footnotesize{}$2.5$} & {\footnotesize{}0.103} & {\footnotesize{}0.543} & {\footnotesize{}0.042}\tabularnewline
\hline 
{\footnotesize{}$3$} & {\footnotesize{}0.020} & {\footnotesize{}0.024} & {\footnotesize{}0.016}\tabularnewline
\hline 
\end{tabular}\tabularnewline
\hline 
\end{tabular}
\end{table}

\section{Case Study: Enron's Dynamic Email Network}

In this section, to show how our proposed method can be applied to
real problems, we model and monitor the Enron email communication
network. The Enron corpus consist of about 0.5 millions of email communications
among $184$ employees of the Enron corporation from 1998 to 2002
\citep{priebe2005scan}. This dataset can be represented by a sequence
of directed networks, where each network represents one week of email
communications. A directed edge is placed between nodes $i$ and $j$
at time $t$ if at least one email has been sent from employee $i$
to the employee $j$ during week $t$. The role of each employee within
the company is available and used to set the attributes of networks.
The roles consist of CEO, president, vice\textendash president, manager,
director, trader, and employee. For simplicity, we focus on the emails
sent among president (P), mangers and directors (MR), and CEO. The
combination of these roles result in a categorical attribute for each
edge with nine possible values (i.e. CEO-CEO, CEO\textendash P, CEO\textendash MR,
...), which are represented by dummy variables. \textcolor{black}{Note
that in these networks, given the attributes (employee roles), the
existence of an edge between two nodes is independent from the existence
of other edges.}

\begin{figure}
\begin{centering}
\includegraphics[width=0.45\columnwidth]{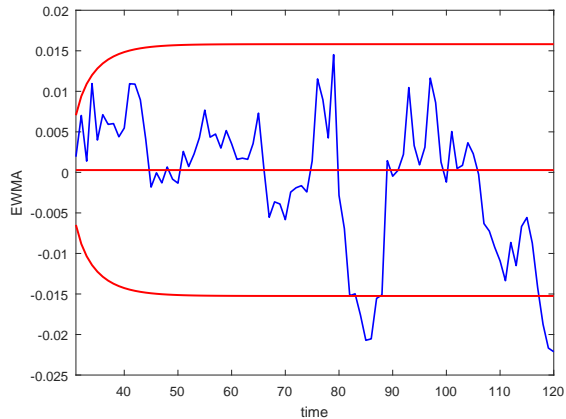}
\par\end{centering}
\caption{EWMA chart of the Pearson residuals obtained from the proposed dynamic
model. The chart signals between week 79 and 89 that matches the time
of a major event in Enron history \label{fig:ewma-enron-1}}
\end{figure}

First, we aggregate 30 initial weeks of data into a single network
and estimate the initial value of the network parameters using the
static method. We select first 30 weeks due to small number of emails
available during that period. We also set $F$ to be identity matrix
and $\xi$ to be zero, which mimics a random walk process. The dynamic
model with EKF is used to infer the parameters of each graph. The
estimated parameters were used to calculate the Pearson residuals
and to detect changes in the stream of networks. We use data from
week 31 to 60 as an in-control stream to estimate the standard deviation
of Pearson residuals used to calculate the control limits. Figure
\ref{fig:ewma-enron-1} illustrates the EWMA chart of the Pearson
residuals obtained from the proposed dynamic model. As can be seen,
the EWMA chart signals between week 79 and 89. These out-of-control
signals relate to the time period that Enron scandal was revealed
\citep{azarnoush2016monitoring}. 

\begin{figure}[H]
\centering{}\subfloat[\label{fig:Estimated-probability-of}]{\begin{centering}
\includegraphics[width=0.45\columnwidth]{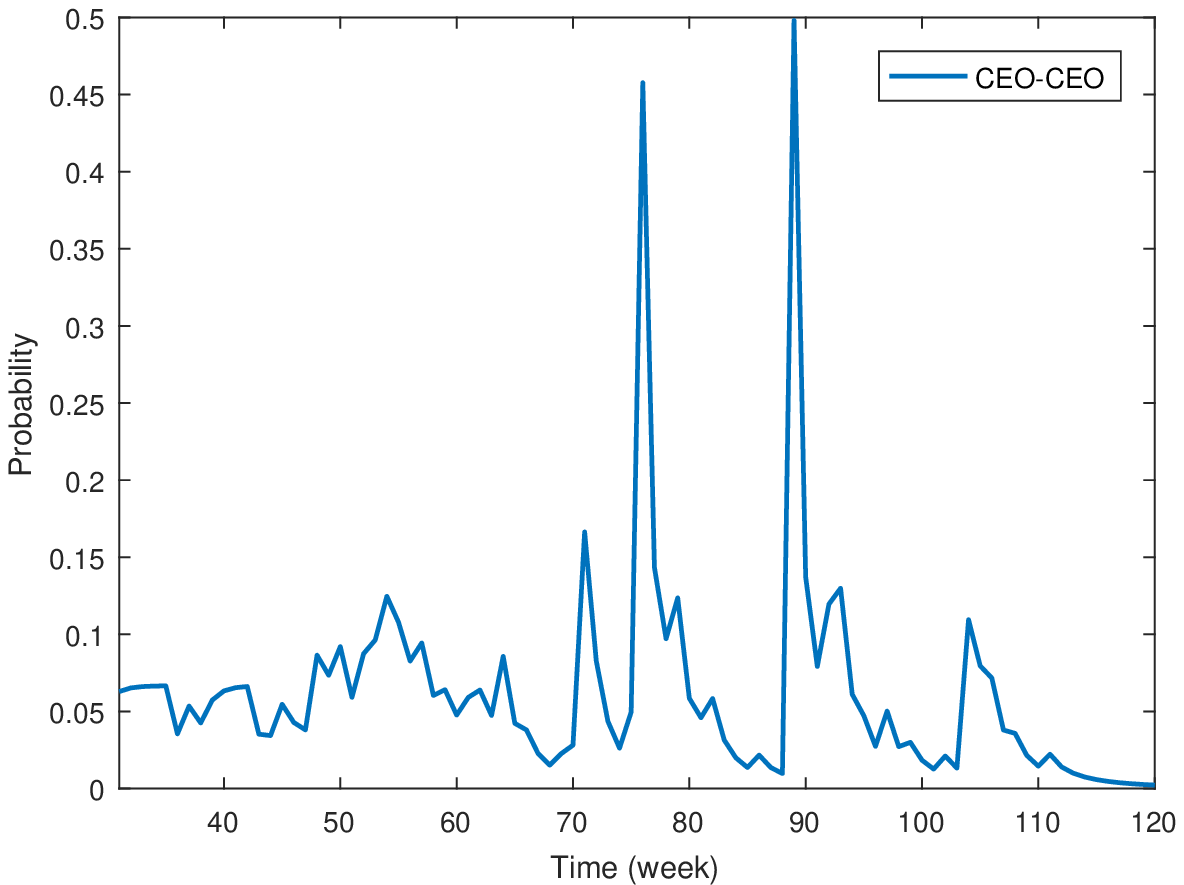}
\par\end{centering}
}\subfloat[\label{fig:ewma-CEOonly}]{\begin{centering}
\includegraphics[width=0.45\columnwidth]{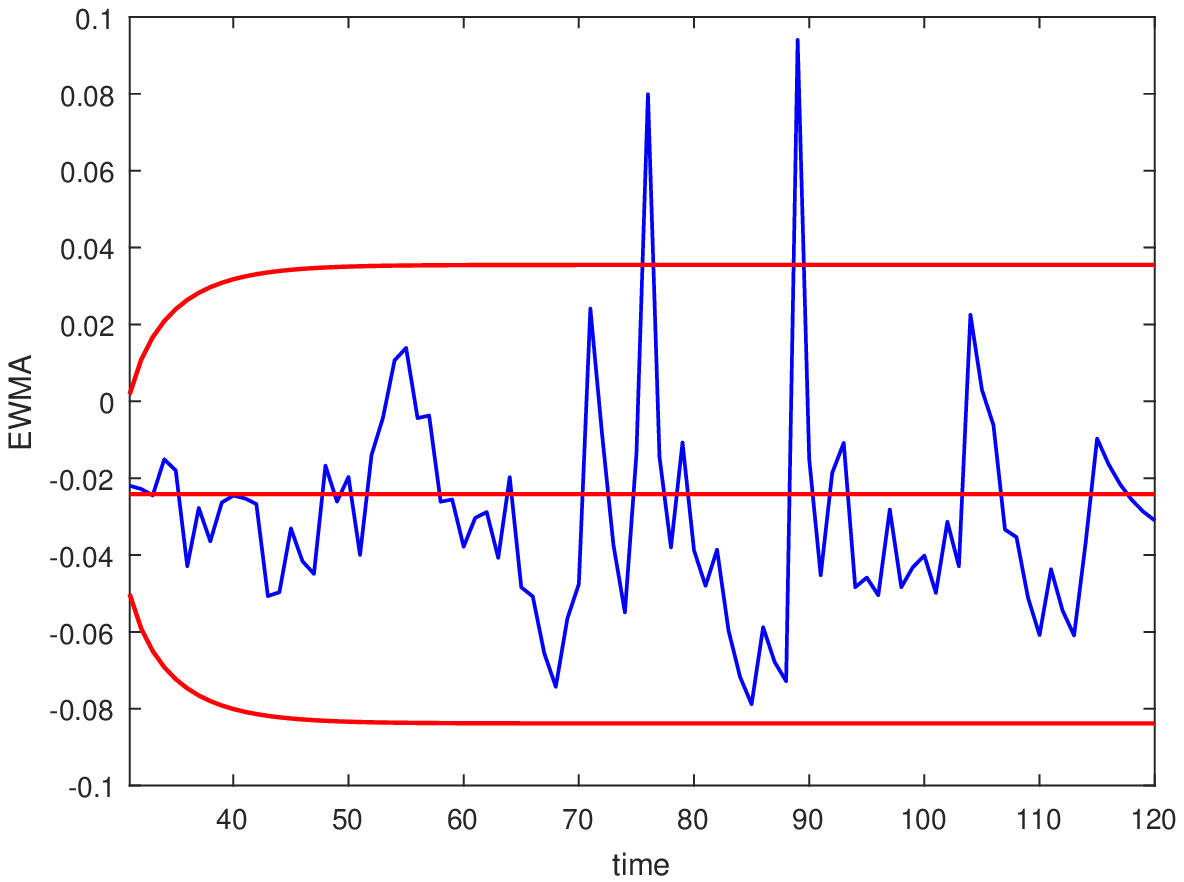}
\par\end{centering}
}\caption{Analysis of Enron data (a) Estimated probability of communication
between two CEOs (b) Change detection using Pearson residuals of the
CEO\textendash CEO sub\textendash graphs.}
\end{figure}

To further explore the change and its potential causes, we calculate
the probability of an edge between two nodes with particular roles.
Because of the few exchanged emails during the first 30 weeks, we
do not include those weeks in the results. Figure \ref{fig:Estimated-probability-of}
illustrates the communication probability of two CEOs. As shown, this
probability drastically changes at weeks $76$ and $89$. This result
is consistent with the observation in the study by \citet{xu2014dynamic}.
Based on this observation, we speculate that the underlying reason
behind the change is an excessive communication among the CEOs. We
setup another EWMA chart, shown in Figure \ref{fig:ewma-CEOonly},
based on the Pearson residual of sub-networks whose nodes represent
the CEOs. As it is illustrated, the control chart detects both jumps
in the CEOs communication. These jumps relate to the CEO resignation
at the Enron Corporation because of the famous scandal. 

\section{Conclusion}

In this paper, we proposed a new methodology for modeling and monitoring
of attributed network streams. First, we modeled static networks using
GLM. Next, we introduced a state transition model over the network
parameters to capture its dynamics. Using the EKF, we predicted network
parameters used for calculating the Pearson residuals (i.e. the standardized
residuals between the predicted and observed networks). Pearson residuals
were then monitored by using EWMA control charts for quick detection
of abrupt changes. We examined the proposed method using both simulated
and real data. In the simulation study, we generated a stream of attributed
networks representing interactions among college students both in
the form of binary and weighted networks. We considered two change
scenarios for each type of the networks, one modeling a global change
and one representing a local change in the stream of networks. The
proposed method was compared with two benchmarks, which fail to fully
capture the network dynamics. The results showed that the proposed
method's detection delay is significantly smaller than two benchmarks,
particularly when the change magnitude is small. Furthermore, the
performance of all three methods deteriorated when detecting local
changes in binary networks. This was due to the lack of data in capturing
the excessive communications. However, the dynamic method still outperformed
both of the benchmarks in detection of local changes. Finally, we
used the Enron corpus as a case study. Our method detected the excessive
communication of the CEO's related to the time period that Enron scandal
was revealed. The results of the case study agree with other studies
on the Enron data. In recent years, several authors proposed different
techniques for community detection in networks. A potential future
research direction is to combine the community detection techniques
with the proposed methodology to detect changes in the community structures
over time. 

\section*{Appendix: Iteratively Reweighted Least Square Method}

Assume that observations $w_{ij,t}$, denoting the weight of the edge
between nodes $i$ and $j$ at time $t$, come from a distribution
in exponential family distribution. That is, $w_{ij,t}\sim f\left(w_{ij,t};\theta_{ij,t}\right)$
with 
\[
f\left(w_{ij,t};\theta_{ij,t}\right)=\exp\left\{ \frac{w_{ij,t}\theta_{ij,t}-b\left(\theta_{ij,t}\right)}{a\left(\phi_{ij,t}\right)}+c\left(w_{ij,t},\phi_{ij,t}\right)\right\} .
\]
Here, $\theta_{ij,t}$ and $\phi_{ij,t}$ are canonical and dispersion
parameters respectively, and $a\left(.\right)$, $b\left(.\right)$,
and $c\left(.\right)$ are known function related to the particular
distribution. We assume $a\left(\phi_{ij,t}\right)=\frac{\phi_{ij,t}}{p}$,
which holds for most of the distributions in the exponential family
including the Bernoulli and Poisson distributions. The constant $p$
is assumed to be known a priori. Now, assume $\theta_{ij,t}=g\left(\left[1,b_{ij}\right]^{\mathtt{T}}\boldsymbol{\beta}_{t}\right)$
where, $g\left(.\right)$ is the link function and $\boldsymbol{x}_{ij}$
is the attribute vector of an edge between nodes $i$ and $j$. 

Given a network with the adjacency matrix $\left[w_{ij,t}\right]$,
and attributes $\boldsymbol{x}_{ij}$, the goal of IRWLS is to estimate
the parameter $\boldsymbol{\beta}_{t}$ by maximizing the log\textendash likelihood
function
\[
l\left(w_{ij,t};\theta_{ij,t}\right)=\sum_{i=1}^{n}\sum_{j=1}^{n}\frac{w_{ij,t}\theta_{ij,t}-b\left(\theta_{ij,t}\right)}{a\left(\phi\right)}+c\left(w_{ij,t},\phi\right)
\]
The IRWLS algorithm works as follows: Given a trial estimate of $\hat{\boldsymbol{\beta}_{t}}$,
one calculates $\eta_{ij,t}=\left[1,\boldsymbol{x}_{ij}\right]^{\mathtt{T}}\hat{\boldsymbol{\beta}_{t}}$,
and $\hat{\theta}_{ij,t}=g\left(\left[1,\boldsymbol{x}_{ij}\right]^{\mathtt{T}}\hat{\boldsymbol{\beta}_{t}}\right)$.
These quantities are used in calculation of working dependent variable
$z_{ij,t}=\hat{\eta}_{ij,t}+\left(w_{ij,t}-\hat{\theta}_{ij,t}\right)\frac{d\eta_{ij,t}}{d\theta_{ij,t}}$,
and the iterative weights $q_{ij,t}=\frac{p}{\left(\frac{\partial^{2}b}{\partial\theta_{ij,t}^{2}}\right)\left(\frac{d\eta_{ij,t}}{d\theta_{ij,t}}\right)^{2}}$,
where all the derivatives are evaluated at the trial estimate. Given
the working dependent variable and the weights, one can re\textendash estimate
the $\boldsymbol{\beta}_{t}$ using the weighted least square formulation
as 
\[
\hat{\boldsymbol{\beta}_{t}}=\left(\boldsymbol{X}^{\mathtt{T}}\boldsymbol{Q}_{t}\boldsymbol{X}\right)^{-1}\boldsymbol{X}^{\mathtt{T}}\boldsymbol{Q}_{t}\boldsymbol{z}_{t},
\]
where $\boldsymbol{X}$ is the matrix of the attributes, $\boldsymbol{Q}_{t}$
is a diagonal matrix with diagonal elements $q_{ij,t}$, and $\boldsymbol{z}_{t}$
is a vector with elements $z_{i}$. This procedure can simply be done
using the $\textrm{Matla\ensuremath{b^{\circledR}}}$ function \textit{glmfit}
from the Statistics and Machine Learning $\textrm{Toolbo\ensuremath{x^{\left(TM\right)}}}$.

\bibliographystyle{authordate4}
\bibliography{Network-Monitoring-refs}

\end{document}